\documentclass[10pt,twocolumn,a4paper]{IEEEtran}
\makeatletter
\def\ps@headings{%
\def\@oddhead{\mbox{}\scriptsize\rightmark \hfil \thepage}%
\def\@evenhead{\scriptsize\thepage \hfil \leftmark\mbox{}}%
\def\@oddfoot{}%
\def\@evenfoot{}}
\makeatother
\IEEEoverridecommandlockouts %/thanks
\usepackage{graphicx,latexsym,subfigure,amsmath,epsfig,latexsym,subfigure,amsmath,cite,amssymb}
\usepackage{amsthm,mathrsfs}
\usepackage{graphicx}
\usepackage{epstopdf}
\usepackage{color}
\usepackage{CJK}
\usepackage{cite} %%
\usepackage{algorithmic}% �㷨
\usepackage{mathrsfs}% ��������
\usepackage{bm}% ϣ����ĸ�Ӻ�
\usepackage{float}% ����ͼƬλ��
\usepackage{setspace}% ʹ�ü����
\usepackage{color}% ʹ����ɫ���
\usepackage{subfigure}% ���벢��ͼƬ
\usepackage{balance}% �������һҳ����
\usepackage{dsfont} %��ʶ����
\usepackage[ruled,linesnumbered]{algorithm2e} % �㷨���̸�ʽ
\SetKwRepeat{DoWhile}{Do}{While}

\begin{document}
\begin{CJK}{GBK}{song}

\title{ Channel Estimation for Underwater Visible Light Communication: A Sparse Learning Perspective}
\author{
        Younan~Mou,
        and Sicong~Liu*,~\IEEEmembership{Senior Member,~IEEE,}
% \thanks{This work was supported in part by the National Natural Science Foundation of China (No. 61901403), in part by the Science and Technology Key Project of Fujian Province, China (No. 2019HZ020009), in part by the Youth Innovation Fund of Natural Science Foundation of Xiamen (No. 3502Z20206039), and in part by the Xiamen Special Fund for Marine and Fishery Development (No. 21CZB011HJ02).(Corresponding Author: Sicong Liu)}
\thanks{Younan Mou and Sicong Liu are with the Department of Information and Communication Engineering, School of Informatics, Xiamen University, Xiamen 361005, China. (\emph{Corresponding Author: Sicong Liu.} email: liusc@xmu.edu.cn).}
\thanks{}
%\author{\small
%\IEEEauthorblockN{Liang Xiao\IEEEauthorrefmark{1},
%Geyi Sheng\IEEEauthorrefmark{1},
%Sicong Liu\IEEEauthorrefmark{1},
%Mugen Peng\IEEEauthorrefmark{2}
%}\\
%\IEEEauthorblockA{\IEEEauthorrefmark{1}Dept. Communication Engineering, Xiamen University, Xiamen 361005, China. Email: \{lxiao; liusc\}@xmu.edu.cn}\\
%\IEEEauthorblockA{\IEEEauthorrefmark{2}Key Laboratory of Universal Wireless Communication (Ministry of Education), Beijing University of Posts and Telecommunications, Beijing,
%China.} %Email: chunguoli@seu.edu.cn
%\IEEEauthorblockA{\IEEEauthorrefmark{3}School of Information Engineering, Xizang Minzu University, Xianyang 712082, China.}%Email: wangcshui@126.com

}
\markboth{}%
{Submitted paper}

\maketitle
\thispagestyle{empty}
\begin{spacing}{1}
\begin{abstract}
  The underwater propagation environment for visible light signals is affected by complex factors such as absorption, shadowing, and reflection, making it very challengeable to achieve effective underwater visible light communication (UVLC) channel estimation.
  It is difficult for the UVLC channel to be sparse represented in the time and frequency domains, which limits the chance of using sparse signal processing techniques to achieve better performance of channel estimation.
  To this end, a compressed sensing (CS) based framework is established in this paper by fully exploiting the sparsity of the underwater visible light channel in the distance domain of the propagation links.
  In order to solve the sparse recovery problem and achieve more accurate UVLC channel estimation, a sparse learning based underwater visible light channel estimation (SL-UVCE) scheme is proposed. Specifically, a deep-unfolding neural network  mimicking the classical iterative sparse recovery algorithm of approximate message passing (AMP) is employed, which decomposes the iterations of AMP into a series of layers with different learnable parameters.
  Compared with the existing non-CS-based and CS-based schemes, the proposed scheme shows better performance of accuracy in channel estimation, especially in severe conditions such as insufficient measurement pilots and large number of multipath components.
\end{abstract}

\vspace{-0.05in}
\section{Introduction}\label{sec:intro}
Realization of high-speed underwater communication system provides an important guarantee for the exploration and development of marine resources.
At present, acoustic waves are widely used in underwater communication systems, but the performance is subject to strong frequency dependent attenuation, resulting in a transmission bandwidth of up to a few hundred kHz \cite{oubei20154}.
In contrast, underwater visible light communication (UVLC) owns hundreds of THz unlicensed bandwidth for communication , and it also has the advantages of high security, low cost and low link delay, etc\cite{pathak2015visible,yang2016priori,yang2016clipping}.
The optical orthogonal frequency division multiplexing (OOFDM) technology has been widely applied in UVLC systems, which effectively mitigates the impact of frequency selective fading and supports high-rate transmission \cite{chen2021joint}.

The study on channel modeling and estimation is among the key directions in the field of UVLC research.
It is critical for the UVLC system to obtain accurate channel state information in order to support high-speed and reliable underwater transmission. However, due to the complex underwater environment, the transmission of light waves is affected by many complex conditions such as absorption, scattering and turbulence \cite{chen2020sherman}, which brings great difficulties to the effectiveness and accuracy of the modelling and estimation of UVLC channels.

In existing research of UVLC channels, the main stream of channel modelling is to obtain the modelling and parameters of the UVLC channel impulse response (CIR) through Monte Carlo numerical simulations \cite{zhang2020monte}.
However, realistic complicated trajectories and states of millions of photons need to be simulated, which costs high computational complexity \cite{zou2022underwater}.
As for the channel estimation methods, a method for UVLC channel estimation is proposed based on inter-symbol frequency-averaging and intra-symbol frequency-averaging \cite{chen2019demonstration}.
However, due to the complex underwater environment, the accuracy of channel estimation is still a great bottleneck limiting the link performance of the UVLC system \cite{chen2020sherman}.
Considering channel estimation for wireless communications, if some prior information such as channel sparsity can be introduced, the compressed sensing (CS) based methods might be utilized to improve the performance.
As for UVLC channel estimation, this becomes a difficult problem because it is difficult for the UVLC channel to be sparse represented in the time and frequency domains as is done for wireless communications.

To this end, in this paper, we extract the inherent sparsity of the UVLC channel frequency response (CFR) and present its sparse representation in the distance domain of the propagation links, thus formulating a novel CS-based framework of UVLC channel estimation.
This makes it possible to use sparse recovery techniques to achieve better performance of channel estimation.
However, the restricted isometry property (RIP), i.e., a coherence metric, of the measurement matrix of the CS-based model for the UVLC channel is not as satisfactory as that of wireless communication channels due to the mechanism of sparse representation in the multipath distance domain.
This might limit the performance of CS-based algorithms in sparse recovery problem solving.

Hence, deep learning techniques widely applied in various fields of wireless communications is introduced to solve the sparse recovery problem with unsatisfactory RIP property.
Specifically, model-driven deep-unfolding neural network models are devised according to the physical characteristics of the system, which can significantly reduce the data requirement and time consumption in training \cite{he2019model,9894087}.
Inspired by this, the classical iterative sparse recovery algorithm of approximate message passing (AMP) \cite{donoho2009message} is decomposed into a series of layers of deep neural networks (DNN) with learnable parameters, and thus a model-driven deep learning method is constructed to realize the estimation of the CFR of the UVLC.

The rest of this paper is structured as follows: The CFR model of the UVLC system is established in Section \uppercase\expandafter{\romannumeral+2};
Section \uppercase\expandafter{\romannumeral+3} formulates the proposed CS-based framework of UVLC channel estimation, and presents the sparse recovery scheme based on model-driven deep-unfolding neural networks;
Section \uppercase\expandafter{\romannumeral+4} reports the simulation results of the proposed scheme with discussions;
Finally, the paper is concluded in Section \uppercase\expandafter{\romannumeral+5}.

\vspace{-0.10in}
\section{System Model}\label{sec:rw}

Considering the important role of UVLC in marine applications, we mainly investigate the visible light channel in the seawater environment in this paper.
Since the optical properties in seawater are only affected by the medium of seawater, the underwater propagation of light waves is affected by the combined effect of the inherent attenuation of pure water and the absorption and scattering of plankton, dissolved substances and suspended particles in the water \cite{lu2021deep}.
Therefore, it can be approximated that the optical properties of seawater are divided into two parts: absorption and scattering \cite{smith1981optical}, resulting in severe attenuation of light waves in seawater.
To sum up, the total attenuation coefficient of the light wave propagation in seawater can be described as

\begin{equation}
  c\left( f \right) = a\left( f \right) + b\left( f \right),\label{1}
\end{equation}
where $a\left( f \right)$ and $b\left( f \right)$ are the absorption and scattering coefficients of seawater, respectively, and their sum $c\left( f \right)$ is the total absorption coefficient of light wave propagation seawater.

\begin{figure}[h]
  \begin{center}
  \vspace{-0.0cm}  %����ͼƬ�����ĵĴ�ֱ����
  \setlength{\abovecaptionskip}{-0.10cm}   %����ͼƬ������ͼ����
  \setlength{\belowcaptionskip}{-0.08cm}   %����ͼƬ���������ľ���
  \includegraphics[width=2.5 in]{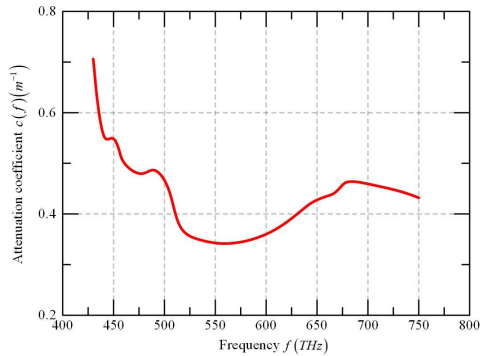}\\
  \caption{Attenuation coefficient with respect to frequency for the UVLC channel \cite{morel2006bio}.}\label{Fig:attenuation}%, $\alpha = 10\%$ and $\sigma^2 = -98.82$ dBm
  \end{center}
\end{figure}

The total attenuation coefficient of light wave propagation in seawater with respect to frequencies \cite{morel2006bio} is shown in Fig \ref{Fig:attenuation}.
It can be shown that the attenuation coefficient of light wave propagation in seawater changes irregularly with frequency, which makes the impact of frequency selectivity more detrimental to the system.
Therefore, the OOFDM technology can be adopted to combat against the severe frequency selective fading of seawater channels.

Note that the available bandwidth of optical spectrum can be up to several hundreds of THz, and the bandwidth actually used is limited by the bandwidth of the transmitter and the receiver, which is usually below of GHz and commonly tens of MHz, Thus, the attenuation coefficient as illustrated in Fig. 1 can be considered approximately regarded as quasi-linear. The channel attenuation coefficient can be modeled as a quasi-linear model $ c\left( f \right) = {c_1}f + {c_2} $ , so the CFR of the underwater wireless optical channel can be rewritten as \cite{kaushal2016underwater}

\begin{equation}
H(f) = \alpha {e^{ - \left( {{c_1}f + {c_2}} \right)s}},
\label{2}
\end{equation}
where $\alpha$ is the attenuation factor of light propagating underwater; $s$ is the distance that light propagates underwater.

\begin{figure}[h]
  \begin{center}
  \vspace{-0.0cm}  %����ͼƬ�����ĵĴ�ֱ����
  \setlength{\abovecaptionskip}{-0.10cm}   %����ͼƬ������ͼ����
  \setlength{\belowcaptionskip}{-0.08cm}   %����ͼƬ���������ľ���
  \includegraphics[width=2.5 in]{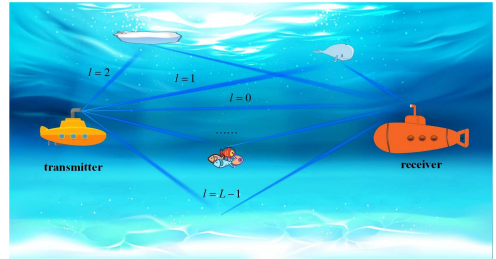}\\
  \caption{The UVLC channel model with a number of paths of visible light propagation.}\label{Fig:model}%, $\alpha = 10\%$ and $\sigma^2 = -98.82$ dBm
  \end{center}
\end{figure}

In addition to the absorption and scattering of the seawater itself, in the complex and changeable marine environment, the UVLC signals can be shielded, reflected or refracted, which leads to the visible light propagation in a number of paths as shown in Fig \ref{Fig:model}.
Therefore, the actual received signal is the superposition of the line-of-sight (LoS) path and multiple non-line-of-sight (NLoS) paths.
Then, the CFR of the underwater visible light channel is rewritten as \cite{ma2018channel}

\begin{equation}
  H\left( f \right) = \sum\limits_{l = 0}^{L -1} {{\alpha _l}{e^{ - \left( {{c_1}f + {c_2}} \right){s_l}}}},
  \label{3}
\end{equation}

where ${\alpha _l}$ and ${s_l}$ are the attenuation factor and propagation distance of the $l$ -th path, respectively; The frequency $f = {f_b} + {f_c}$ is composed of the baseband signal frequency ${f_b}$ and the optical carrier frequency ${f_c}$ ; $L$ represents the number of LoS and NLoS paths.

A number of pilot subcarriers with equal spacing are employed to estimate the CFR of the UVLC system.
Assuming that $P$ uniformly spaced pilot subcarriers are used, the pilot frequency ranges from ${f_{\min }}$ to ${f_{\max }}$ with the pilot frequency interval of $\Delta f = \left( {{f_{\max }} - {f_{\min }}} \right)/\left( {P - 1} \right)$ .
Thus, the frequency of each pilot subcarrier is given by ${f_i} = {f_{\min }} + i\Delta f$ , $1 < i < P$ .
Then, the received $i$ -th pilot subcarrier can be represented by

\begin{equation}
  H\left( {{f_i}} \right) = \sum\limits_{l = 0}^{L - 1} {{\alpha _l}{e^{ - \left[ {{c_1}\left( {{f_{\min }} + i\Delta f} \right) + {c_2}} \right]{s_l}}}},\label{4}
\end{equation}
the received pilots $\{ H\left( {{f_i}} \right)\} _{i = 1}^P$ can be regarded as the measurement data of the CFR.

%\vspace{-0.15in}
\section{Channel Estimation Scheme For Underwater Visible Light Communication Based On Sparse Learning}\label{sec:sm}

In order to make better use of the sparsity of the CFR in the distance domain, Eq. \ref{4} is rewritten as

\begin{equation}
  H\left( {{f_i}} \right) = \sum\limits_{l = 0}^{L - 1} {{\alpha _l}{e^{ - \left[ {{c_1}\left( {{f_{\min }} + i\Delta f} \right) + {c_2}} \right]{s_l}}}}  = \sum\limits_{l = 0}^{L - 1} {\omega _l^i \cdot {\lambda _l}},\label{5}
\end{equation}
where we define some channel parameters as ${\omega _l} \buildrel \Delta \over = {e^{ - {c_1}\Delta f{s_l}}}$ and ${\lambda _l} \buildrel \Delta \over = {\alpha _l}{e^{ - \left( {{c_1}{f_{\min }} + {c_2}} \right){s_l}}}$ for the convenience of modelling and representation.
Note that $\left\{ {{\lambda _l},0 \le l \le {L - 1}} \right\}$ is only related to the propagation distance ${s_l}$ of a specific UVLC path.
When the value of the channel parameters $\left\{ {{\lambda _l},0 \le l \le {L - 1}} \right\}$ and the path distance ${s_l}$ is obtained, the value of the attenuation factor ${\alpha _l}$ can be calculated.

Considering that in the actual seawater environment, the number of UVLC paths $L$ between the transmitter and receiver is limited, the distances of all the $L$ paths $\left\{ {{s_l}} \right\}_{l = 0}^{L-1}$ are only a few values selected out of the distance domain, leading to the sparsity of the CFR in the distance domain, which is explained later in detail.
To facilitate the sparse representation of the distance domain, we can discretize the continuous distance ${s_l}$ into discrete values, i.e., ${n_l}\Delta s$ , where ${n_l}$ is the quantization coefficient, and $\Delta s$ is the minimum quantization step size of the path.
In practical application, a proper step size of $\Delta s$ can be determined to make a good tradeoff between computational complexity and channel estimation accuracy.
Increasing the step size can reduce the computational complexity, while it will also lead to a degradation in the accuracy of channel estimation.

In the UVLC system, we can formulate the channel measurement vector using the received pilots, which is actually the sampling of the CFR as given by

\begin{equation}
  {\textbf{y}} = \left[ {H\left( {{f_1}} \right),H\left( {{f_2}} \right), \cdots ,H\left( {{f_{P}}} \right)} \right].\label{6}
\end{equation}

At this point, the channel estimation problem is transferred into a parametric sparse recovery problem, which can be solved effectively using CS-based methods by fully exploiting the sparse characteristics of the CFR in the distance domain.
Therefore, a CS-based UVLC channel measurement model is established as

\begin{equation}
  {\textbf{y} = \boldsymbol{\Phi}\textbf{x} + \textbf{w}},\label{7}
\end{equation}
where the observation matrix $\boldsymbol{\Phi} \in \mathbb{C}^{P \times N}$ is expressed as

\begin{equation}
  {\boldsymbol{\Phi }} = \left[ {\begin{array}{*{20}{c}}
    {{v_0}}&{{v_1}}& \cdots &{{v_{N - 1}}}\\
    {v_0^2}&{v_1^2}& \cdots &{v_{N - 1}^2}\\
     \vdots & \vdots & \ddots & \vdots \\
    {v_0^P}&{v_1^P}& \cdots &{v_{N - 1}^P}
    \end{array}} \right],\label{8}
\end{equation}
where ${v_k} = {e^{ - {c_1}\Delta f\Delta s \cdot k}}$ is the quantized expression of the parameter ${\omega _l}$ ;
\textbf{w} is the background noise vector;
The unknown vector ${\textbf{x}} = {\left[ {{x_0},{x_1}, \cdots ,{x_{N - 1}}} \right]^T}$ is a sparse vector containing $L$ non-zero elements.
For the convenience of expression, ${\textbf{x}}$ is called the channel sparse proxy vector.
The subscripts of each element of ${\textbf{x}}$ ranging from $0$ to $N-1$ are related to the distance from $0$ to $\left( {N - 1} \right)\Delta s$ in the distance domain, respectively.
The length $N$ of the vector \textbf{x} is determined by the NLoS path with the maximum distance ${s_{\max }}$ in the UVLC channel, i.e., $N = \left\lceil {{s_{\max }}/\Delta s} \right\rceil $.

For any element ${x_k}$ in the channel sparse proxy vector \textbf{x} , if there exists a path with the distance of ${s_l} = k \cdot \Delta s$ in the UVLC channel, then the value of ${x_k}$ is the distance-quantized expression of ${\lambda _l}$ , i.e., ${x_k} = {\alpha _l}{e^{ - \left( {{c_1}{f_{\min }} + {c_2}} \right)k \cdot \Delta s}}$ ;
Otherwise, if the path does not exist, the corresponding element is simply zero, i.e., ${x_k} = 0$ .
In the UVLC channel environment, the actual number of paths $L$ between the transmitter and receiver is limited, so the distances of all the $L$ paths $\left\{ {{s_l}} \right\}_{l = 0}^{L-1}$ are only a few values selected out of the discretized distance domain, i.e., $L \ll N$ .
Thus, the channel sparse proxy vector \textbf{x} is a sparse vector reflecting the sparse characteristics of the CFR in the distance domain.
The distance of each path is corresponding to a unique element of the channel proxy sparse vector \textbf{x} .
Then, the support $\Omega $ of the sparse vector \textbf{x} , i.e., the set of the subscripts of the non-zero elements, is given by

\begin{equation}
  \Omega  = \left\{ {\left\lfloor {{s_l}/\Delta s} \right\rfloor \left| {l = 0, \cdots ,L - 1} \right.} \right\}.\label{9}
\end{equation}

As formulated in the CS-based UVLC channel measurement model in Eq. \ref{7} , the channel measurement vector \textbf{y} has been represented as the product of an observation matrix and an unknown sparse vector \textbf{x} .
Note that the observation matrix $\boldsymbol{\Phi}$ is much underdetermined, i.e., the number of columns of the observation matrix is much larger than the number of rows.
This means that the amount of measurement data is much less than that of the size of the unknown vector to be reconstructed.

The performance of the CS-based methods for sparse recovery is very much related to the mutual incoherence property (MIP) of the observation matrix $\boldsymbol{\Phi}$, which is a sufficient condition for the restricted isometric property (RIP) \cite{ben2010coherence}.
RIP can be also regarded as a metric to indicate the incoherence of the observation matrix, which is an important constraint affecting the performance of CS-based methods \cite{donoho2006compressed}.
Since it is easier for the MIP to be quantitatively evaluated, it is used to measure the incoherence of the observation matrix in this paper.
Specifically, the MIP of the observation matrix $\boldsymbol{\Phi}$ of the CS-based UVLC channel measurement model in Eq. \ref{7} is defined as

\begin{figure*}[t]
  \begin{center}
  \vspace{-0.0cm}
  \setlength{\abovecaptionskip}{-0.10cm}
  \setlength{\belowcaptionskip}{-0.08cm}
  \includegraphics[width=6 in]{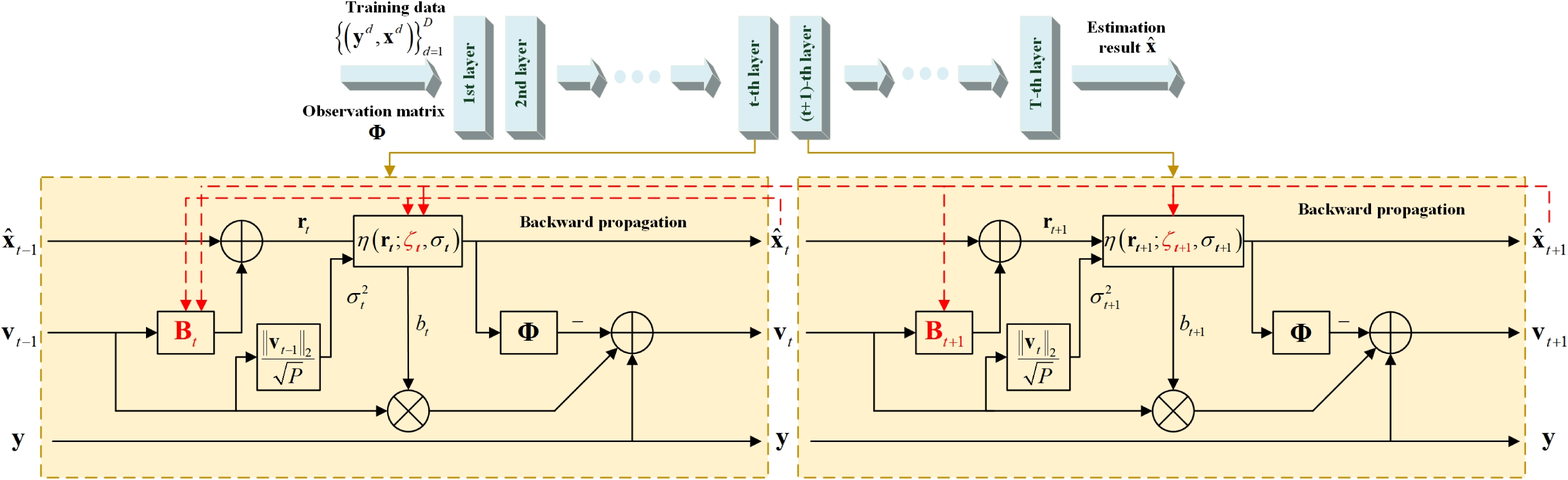}\\
  \caption{Sparse-learning-based framework of UVLC channel estimation: A deep unfolded neural network mimicking the iterations of the AMP algorithm is exploited; Parameters of each layer of the network can be learnt in training; Channel estimation in practice can be implemented by simple forward propagation using trained model.}\label{Fig:LAMP}
  \end{center}
\end{figure*}

\begin{equation}
  \mu  = \mathop {\max }\limits_{0 \le j,k \le N - 1,j \ne k} \frac{{\left| {\left\langle {{\varphi _j},{\varphi _k}} \right\rangle } \right|}}{{{{\left\| {{\varphi _j}} \right\|}_2} \cdot {{\left\| {{\varphi _k}} \right\|}_2}}},\label{10}
\end{equation}
where ${\varphi _j}$ and ${\varphi _k}$ represent the $j$ -th and $k$ -th columns of the observation matrix $\boldsymbol{\Phi}$ , respectively;
$\left\langle {{\varphi _j},{\varphi _k}} \right\rangle $ is the inner-product of the two column vectors.

According to Eq. \ref{10}, it can be derived that the MIP of the observation matrix given in Eq. \ref{8} is generally not small, and even might be close to one in some extreme case, resulting in weak RIP.
In this case, the classical CS-based methods for sparse recovery will suffer from significant performance degradation, or even do not work \cite{ma2018channel}.
To this end, we utilize the deep learning technology to enhance the sparse recovery methods for sparse UVLC channel estimation.

Specifically, a deep sparse-learning-based framework of UVLC channel estimation is formulated as shown in Fig. \ref{Fig:LAMP}.
The classical iterative sparse approximation algorithm of AMP is mimicked by a deep-unfolding neural network.
The parameters of each layer of the network can be learnt in training.  Considering the bottleneck of weak RIP of the observation matrix, we need to relieve the constraint on the incoherence of the observation matrix, and find another approach to compensate for the capability of sparse recovery.
Through the data-driven approach in deep learning, we can obtain some inherent features out of the UVLC channel apart from sparsity \cite{machidon2022deep}.
In this way, the performance of deep sparse learning enabled channel estimation in UVLC environments can be significantly improved.

The proposed scheme of sparse learning based UVLC channel estimation (SL-UVLC) consists of two phases, i.e., the training phase and inference phase, which is summarized in \textbf{Algorithm 1}.

In \textbf{Algorithm 1}, $\hat{\textbf{x}}_{t}$ is the channel sparse proxy vector recovered by the $t$ -th layer; ${{\textbf{v}}_t}$ is the residual of the $t$ -th layer, and ${b_t}{{\textbf{v}}_{t - 1}}$ is called the $Onsager$  $Correction$ term;
${\sigma _t}$ is the standard deviation estimate of ${{\textbf{v}}_t}$ ;
${{\textbf{B}}_t}$ and ${\zeta _t}$ are the learnable parameters that vary with the number of layers, where ${{\textbf{B}}_t}$ is initialized by the observation matrix, i.e., ${{\textbf{B}}_t} \buildrel \Delta \over = {{\boldsymbol{\Phi }}^H}$ , and ${\zeta _t}$ is used to control the fidelity of the soft threshold function $\eta(\bullet ; \bullet)$ , which is expressed as

\begin{equation}
  \begin{aligned}
  {\left[\eta\left(\textbf{r}_t ; \zeta_t, \sigma_t\right)\right]_i } &=\eta\left(\left|r_{t, i}\right| ; \zeta_t, \sigma_t\right) \\
  &=\operatorname{sgn}\left(r_{t, i}\right) \max \left(\left|r_{t, i}\right|-\zeta_t \sigma_t, 0\right).
  \end{aligned}
  \label{11}
\end{equation}
The loss function ${L_t}\left( {\boldsymbol{\Theta }} \right)$ of the $t$ -th layer is  given by

\begin{equation}
  L_t(\boldsymbol{\Theta})=\frac{1}{D} \sum_{d=1}^D\left\|\hat{\textbf{x}}_{T}\left(\textbf{y}^{(d)} ; \boldsymbol{\Theta}\right)-\textbf{x}^{(d)}\right\|_2^2,\label{12}
\end{equation}
where ${{\textbf{x}}^{\left( d \right)}}$ and ${{\textbf{y}}^{\left( d \right)}}$ represent the ground-truth channel sparse proxy vector and the channel measurement vector of the $d$ -th training data sample, respectively, and $\hat{\textbf{x}}_{T}\left( {{{\textbf{y}}^{\left( d \right)}};{\boldsymbol{\Theta }}} \right)$ is the estimated channel sparse proxy vector which is the output of the network.

\begin{algorithm}[t]
\caption{Sparse Learning Based Underwater Visible Light Channel Estimation (SL-UVCE).}
\tcp{\bf{Training Phase}}
\KwIn{\\ 1) The training set $\left\{ {\left( {{{\textbf{y}}^{d}},{{\textbf{x}}^{d}}} \right)} \right\}_{d = 1}^D$ consisting of the channel measurement vector ${{\textbf{y}}^{d}}$ and the ground-truth channel sparse proxy vector ${{\textbf{x}}^{d}}$ ; \\ 2) Observation matrix ${\boldsymbol{\Phi }}$ }
Initialization: $P$ ; $N$ ; ${{\textbf{v}}_0} \leftarrow {\textbf{0}}$ ; ${b_0} \leftarrow 0$ ; ${hat{\textbf {x}}_{0}} \leftarrow {\textbf{0}}$ ;\\
\For{$t = 1, 2, 3,...$}{
  Initialize the learnable parameters: ${{\textbf{B}}_t} \leftarrow {{\boldsymbol{\Phi }}^H}$ ; ${\zeta _t} \leftarrow 1$\\
  ${{\textbf{v}}_t} = {\textbf{y}} - {\boldsymbol{\Phi }}{\hat{\textbf{x}}_{t - 1}} + {b_t}{{\textbf{v}}_{t - 1}}$\\
  $\sigma _t^2 = \frac{1}{P}\left\| {{{\textbf{v}}_t}} \right\|_2^2$\\
  ${{\textbf{r}}_t} = {{\textbf{x}}_{t - 1}} + {{\textbf{B}}_t}{{\textbf{v}}_t}$\\
  ${\hat{\textbf{x}}_t} = \eta \left( {{{\textbf{r}}_t};{\zeta _t},{\sigma _t}} \right)$\\
  ${b_t} = \frac{1}{P}{\left\| {\hat{\textbf{x}}_t} \right\|_0}$\\
  Calculate the loss function ${L_t}\left( {\boldsymbol{\Theta }} \right)$ using Eq. \ref{12}\\
  Update learnable parameters ${{\boldsymbol{\Theta }}_t} = \left\{ {{{\textbf{B}}_t},{\zeta _t}} \right\}$\\
  \textbf{If} ${L_t}\left( {\boldsymbol{\Theta }} \right) > {L_{t - 1}}\left( {\boldsymbol{\Theta }} \right)$ , set the number of layers $T$ to $t-1$ , \textbf{break}.}
  \KwOut{\\ 1) Learned parameters ${\boldsymbol{\Theta }} = \left\{ {{{\textbf{B}}_t},{\zeta _t}} \right\}_{t = 1}^T$\\ 2)	The final number of layers $T$ }
%  \vspace{10 pt}
  \tcp{\bf{Inference Stage}}
  \KwIn{\\ 1) Realistic channel measurement vector ${\textbf{y}}$ \\ 2)	Learned parameters ${\boldsymbol{\Theta }} = \left\{ {{{\textbf{B}}_t},{\zeta _t}} \right\}_{t = 1}^T$ of the trained network}
  Initialization: $\hat{\textbf{x}} \leftarrow {\textbf{0}}$ \\
  Estimate the sparse proxy vector $\hat{\textbf{x}} = \eta \left( {{{\textbf{r}}_T};{\zeta _T},{\sigma _T}} \right)$ through feed-forward propagation in the trained network\\
  \KwOut{\\ Estimated sparse proxy vector $\hat{\textbf{x}}$ }
\end{algorithm}

In the training phase of {\textbf{Algorithm 1}}, using the UVLC channel training dataset $\left\{ {\left( {{{\textbf{y}}^d},{{\textbf{x}}^d}} \right)} \right\}_{d = 1}^D$ , the learnable network parameters ${\boldsymbol{\Theta }} = \left\{ {{{\textbf{B}}_{t}},{\zeta _t}} \right\}_{t = 1}^T$ of the sparse learning based model-driven deep-unfolding networks are trained layer-wise to extract the inherent features of the UVLC channel.
When optimizing the $t$-th layer, the network parameters learned by the previous layer is fixed, and the learnable parameters ${\boldsymbol{\Theta }} = \left\{ {{{\boldsymbol{B}}_t},{\zeta _t}} \right\}_{t = 1}^T$ of the current $t$ -th layer is trained.
The loss function ${L_t}\left( {\boldsymbol{\Theta }} \right)$ is jointly optimized by back propagation and the stochastic gradient descent.
If ${L_t}\left( {\boldsymbol{\Theta }} \right) < {L_{t - 1}}\left( {\boldsymbol{\Theta }} \right)$ , the number of network layers is increased by one and the training is continued to the next layer;
Otherwise if ${L_t}\left( {\boldsymbol{\Theta }} \right) > {L_{t - 1}}\left( {\boldsymbol{\Theta }} \right)$ , the training terminates and the final number of layers is set as $T = t - 1$ .

In the inference phase of {\textbf{Algorithm 1}}, the realistic channel measurement vector \textbf{y} are input into the trained network, and a single feed-forward operation can be used to efficiently estimate the sparse proxy vector $\hat{\textbf{x}}$ .
Then, the channel parameters $\left\{ {{\lambda _l},0 \le l < L} \right\}$ , the propagation distance ${s_l}$ and the attenuation factor ${\alpha _l}$ can all be obtained from $\hat{\textbf{x}}$ according to Eq. \ref{5}, which realizes the estimation  of the UVLC channel.

\section{Simulation Results}\label{sec:learn}

In this section, the performance of the proposed SL-UVCE scheme for UVLC channel estimation based on sparse learning is evaluated by simulations with respect to the amount of available measurement pilots and the number of paths of the UVLC channel.
The existing non-CS-based least squares (LS) scheme \cite{shi2020adaptive} and the classical CS-based algorithm of orthogonal matching pursuit (OMP) \cite{niaz2016compressed} are also evaluated for comparison.
The simulation parameters used for UVLC channel estimation are listed in Table \uppercase\expandafter{\romannumeral1}.

\begin{table}[h]
  \caption{Simulation Parameters}\label{Tab:Simulation Parameters}
  \centering
  \begin{tabular}{llr}
  \hline
  CP length   &  & 256                  \\ \hline
  OFDM symbol size   &  & 2048                  \\ \hline
  UVLC central frequency   &  & 16MHz      \\ \hline
  UVLC bandwidth       &  & 28MHz      \\ \hline
  Light wavelength &  & 450nm       \\ \hline
  Underwater optical ambient noise &  & AWGN       \\ \hline
  \end{tabular}
\end{table}

For the training phase, randomly generate the training set $\left\{ {\left( {{{\textbf{y}}^d},{{\textbf{x}}^d}} \right)} \right\}_{d = 1}^D$ of size $D = 1000$ according to the system model in Section \uppercase\expandafter{\romannumeral+2}, and generate the testing set $\left\{ {\left( {{{\textbf{y}}^s},{{\textbf{x}}^s}} \right)} \right\}_{s = 1}^S$ of size $S = 100$ in a similar way to prevent overfitting of the deep-unfolding neural network. The Adam optimizer is adopted to optimize the learnable network parameters ${\boldsymbol{\Theta }} = \left\{ {{{\textbf{B}}_t},{\zeta _t}} \right\}_{t = 1}^T$ with the learning rate ${l_r} = 0.01$ .
The performance of UVLC channel estimation is evaluated by normalized mean square error (NMSE) defined as

\begin{equation}
  \mathrm{NMSE}=\mathbb{E}\left[\frac{\|\textbf{x}-\hat{\textbf{x}}\|_2^2}{\|\textbf{x}\|_2^2}\right],\label{13}
\end{equation}
where $\hat{\textbf{x}}$ represents the estimated channel sparse proxy vector result ; $\|\bullet\|_2$ and $\mathbb{E}[\bullet]$ denotes the ${l_2}$ -norm and the expectation operator, respectively.

\begin{figure}[h]
  \begin{center}
  \vspace{-0.0cm}  %����ͼƬ�����ĵĴ�ֱ����
  \setlength{\abovecaptionskip}{-0.10cm}   %����ͼƬ������ͼ����
  \setlength{\belowcaptionskip}{-0.08cm}   %����ͼƬ���������ľ���
  \includegraphics[width=3.5 in]{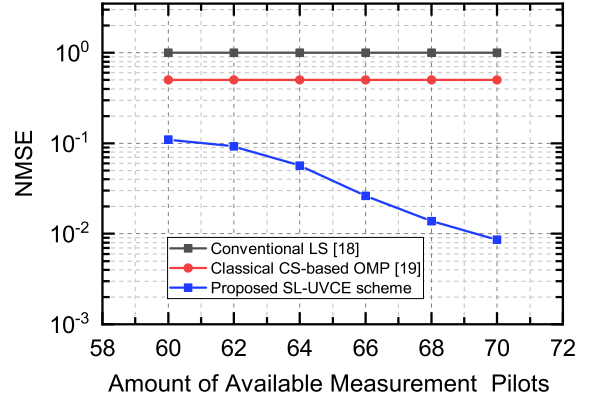}\\
  \caption{The performance of UVLC channel estimation  underwith respect to different the amount of available measurement  pilot numbers }\label{Fig:Pilots}%, $\alpha = 10\%$ and $\sigma^2 = -98.82$ dBm
  \end{center}
\end{figure}

The performance of the proposed scheme of UVLC channel estimation with respect to the amount of available measurement pilots is reported in Fig. \ref{Fig:Pilots}.
It is observed from Fig. \ref{Fig:Pilots} that the proposed SL-UVCE scheme achieves satisfactory performance in estimating the realistic UVLC channel.
It is shown that the NMSE decreases with the amount of available measurement pilots for the proposed scheme. With less than $70$ pilots, the NMSE of the proposed scheme can reach below ${10^{ - 2}}$ , which can already support reliable underwater wireless broadband transmission in case of severe channel conditions like offshore areas with relatively large number of channel paths. In comparison, it can be observed from Fig. \ref{Fig:Pilots} that the performance of both the classical CS-based algorithm of OMP and the conventional non-CS-based LS scheme hardly improve with the pilot amount, which validates the effectiveness of the proposed scheme in estimating the complex UVLC channel.
It is also observed that the CS-based OMP algorithm outperforms the LS scheme, because the sparse characteristics of the UVCL channel is utilized.
However, due to the weak RIP condition of the observation matrix of the CS-based UVLC channel estimation model, the accuracy of the sparse recovery of the channel is limited. On the contrary, the proposed SL-UVCE scheme has introduced the deep-unfolding neural network to better extract the sparse features of the UVLC channel and to learn the inherent channel characteristics other than sparseness.
Therefore, in case of insufficient pilots, the proposed scheme can achieve significant performance improvement compared with the benchmark schemes.

\begin{figure}[h]
  \begin{center}
  \vspace{-0.0cm}  %����ͼƬ�����ĵĴ�ֱ����
  \setlength{\abovecaptionskip}{-0.10cm}   %����ͼƬ������ͼ����
  \setlength{\belowcaptionskip}{-0.08cm}   %����ͼƬ���������ľ���
  \includegraphics[width=3.5 in]{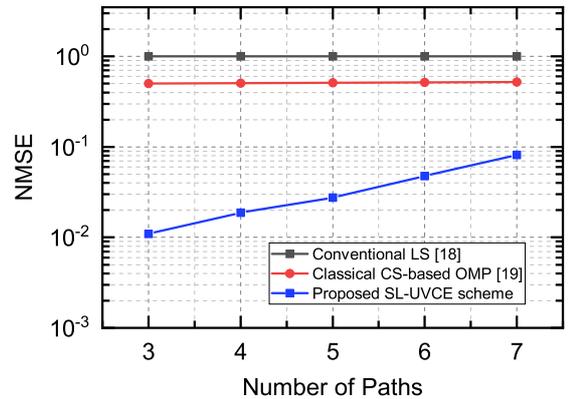}\\
  \caption{The performance of UVLC channel estimation with respect to the number of channel paths}\label{Fig:paths}%, $\alpha = 10\%$ and $\sigma^2 = -98.82$ dBm
  \end{center}
\end{figure}

The performance of the proposed scheme of UVLC channel estimation with respect to the number of channel paths is reported in Fig. \ref{Fig:paths}.
It is shown from the simulation results in Fig. \ref{Fig:paths} that, the proposed SL-UVCE scheme can achieve satisfactory performance of UVLC channel estimation in condition of different number of channel paths.
Even in severe underwater conditions such as offshore areas with a relatively large number of channel paths, the proposed scheme can also reach high accuracy of channel estimation.
It can also be noted from Fig. \ref{Fig:paths} that, the proposed scheme significantly outperforms the CS-based and non-CS-based benchmark schemes, which verifies the capability of the proposed scheme in the sparse feature extraction from the UVLC channel and the sparse recovery of CS-based measurement model with weak RIP for severe seawater environments.

%\vspace{-0.1in}
\section{Conclusion}\label{sec:conclusion}

In this paper, a compressed sensing (CS)-based framework of underwater visible light communication (UVLC) channel estimation has been established to exploit the sparsity of the channel frequency response (CFR) of the UVLC in the distance domain.
A scheme of sparse learning based underwater visible light channel estimation (SL-UVCE) has been proposed to solve the formulated sparse recovery problem.
The inherent sparsity of the CFR in the distance domain corresponding to multiple channel paths has been fully utilized to formulate the sparse representation of the UVLC channel.
Moreover, in order to overcome the problem of weak RIP of the CS-based UVLC measurement model, a model-driven deep-unfolding neural network mimicking the iterations of the iterative sparse approximation algorithm has been devised, which can learn deeper features other than sparsity of the UVLC channel.
Simulation results have verified the accuracy and effectiveness of the proposed SL-UVCE scheme, which significantly outperforms the existing CS-based and conventional non-CS-based benchmark schemes, especially in complex conditions, such as insufficient available pilots or severe seawater environments with more channel paths.
It is promising for the proposed scheme to be applied in realistic complex environments like offshore and shallow sea waters to satisfy the requirements of effective and reliable broadband underwater transmission.

%\vspace{-0.15in}
\bibliography{VLC_PU}
\bibliographystyle{IEEEtr}

%\balance
\end{spacing}
\end{CJK}
\end{document}